\documentclass[structabstract]{aa}  
\usepackage{graphicx}
\usepackage{txfonts}
\newcommand{\reduceme}{\mbox{R\raisebox{-0.35ex}{E}D%
\hspace{-0.05em}\raisebox{0.85ex}{uc}\hspace{-0.90em}%
\raisebox{-.35ex}{{m}}\hspace{0.05em}E}}

\begin{document}

\title{Evidence of a massive planet candidate\\
orbiting the young active 
K5V star BD+20 1790
\thanks{Based on observations collected at the German-Spanish
  Astronomical Center, Calar Alto, jointly operated by the
  Max-Planck-Institut f\"ur Astronomie Heidelberg and the Instituto de
  Astrof\'isica de Andaluc\'ia (CSIC). Based on observations made with
  the Italian Telescopio Nazionale Galileo (TNG) operated on the
  island of La Palma by the Fundaci\'on Galileo Galilei of the INAF
  (Istituto Nazionale di Astrofisica) at the Spanish Observatorio del
  Roque de los Muchachos of the Instituto de Astrof\'isica de
  Canarias. Based on observations made with the Liverpool Telescope
  operated on the island of La Palma by Liverpool John Moores
  University in the Spanish Observatorio del Roque de los Muchachos of
  the Instituto de Astrof\'isica de Canarias with financial support
  from the UK Science and Technology Facilities Council.}}

   \author{M. Hern\'an-Obispo
          \inst{1}
          \and M.C. G\'alvez-Ortiz
          \inst{2}
          \and G. Anglada-Escud\'e
          \inst{3,4}
          \and S.R. Kane
          \inst{5}
          \and J.R. Barnes
          \inst{2}
          \and E. de Castro 
          \inst{1}
          \and M. Cornide
          \inst{1}
          }

   \institute{Dpto. de Astrof\'isica y Ciencias de la
  Atm\'osfera, Facultad de F\'isica, Universidad Complutense
  de Madrid, Avda. Complutense s/n, E-28040, Madrid, Spain 
              \\
              \email{mho@astrax.fis.ucm.es}
         \and
             Centre for Astrophysics Research, Science \& Technology Research Institute,\\ University of Hertfordshire, College Lane, Hatfield, Hertfordshire AL10 9AB, UK
          \and 
	Department of Terrestrial Magnetism, Carnegie Institution of Washington, 5241 Broad Branch Road, NW, Washington,
  DC 20015-1305, USA
           \and 
      Departament d $^\prime$Astronomia i Meteorologia.
Universitat de Barcelona, Mart\'i i Franqu\'es 1, Barcelona, 08028. Spain
	   \and
NASA Exoplanet Science Institute, Caltech, MS
  100-22, 770 South Wilson Avenue, Pasadena, CA 91125, USA
             }

   \date{recieved -- ; accepted --}

\abstract{BD+20 1790 is a young active, metal-rich, late-type K5Ve
  star. We have undertaken a study of stellar activity and
  kinematics for this star over the past few years. Previous results
  show a high level of stellar activity, with the presence of
  prominence-like structures, spots on the surface and strong flare
  events, despite the moderate rotational velocity of the star. In addition,
  radial velocity variations with a semi-amplitude of up to 1 km
  s$^{-1}$ were detected.}
% aims
{We investigated the nature of these radial velocity variations, in
  order to determine whether they are due to stellar activity or
  the reflex motion of the star induced by a companion.}
% methods
{We have analysed high-resolution echelle spectra, by measuring
  stellar activity indicators, and computing radial velocity (RV) and
  bisector velocity spans. Also two-band photometry was obtained to
  produce the light curve and determine the photometric period.  }
% results
{Based upon the analysis of the bisector velocity span, as well as
  spectroscopic indices of chromospheric indicators, like e.g. Ca~{\sc
    II} H \& K, H$\alpha$, and taking into account the photometric
  analysis, we report that the best explanation for the RV variation
  is the presence of a sub-stellar companion. The Keplerian fit of the
  RV data yields a solution for a close-in massive planet with an
  orbital period of 7.78 days. The presence of the close-in massive
  planet could also be an interpretation for the high level of stellar
  activity detected. Since the RV data are not part of a planet search
  program, we can consider our results as a serendipitous evidence of  a planetary companion. To date, this is the youngest main sequence
  star for which a planetary candidate has been reported.  }{}

\titlerunning{Massive candidate orbiting young K5 star}

   \maketitle
%
%________________________________________________________________

\section{Introduction}

Since the detection of the first planet orbiting a main sequence star,
51 Peg (Mayor \& Queloz 1995), the radial
velocity (RV) method has become the most successful technique for
detecting exoplanets as the vast majority have thus far been
discovered in this way (Udry \& Santos 2007). This method is
especially efficient for giant planets in close-in orbits owing to the large radial velocities they induce in the host star. The use of
the RV technique to detect exoplanets around young and active stars
requires, in addition, a careful characterization of stellar
activity. An active region on the stellar surface can produce changes
in the shape of the spectral lines, thus inducing a subsequent
temporal variation of the RVs that may mimic a planetary reflex motion
with a period equal to the rotational period of the star (Saar \&
Donahue 1997). Some cases of false planetary detections are provided
by Queloz et al. (2001), Bouvier et al. (2007), Huerta et al. (2008) and
Hu\'elamo et al. (2008).
Thus the challenge in using the RV technique to detect young planets
lies in disentangling the increased levels of stellar activity of
young stars from the RV signals of the planets.

There is an absence of planets detected around stars younger than 100
Myr (Setiawan et al. 2007, Setiawan et al. 2008). Most RV searches for
planetary companions have focussed mainly on stars older than 1
Gyr. Young stars were omitted from RV surveys until recently. Nevertheless, great effort has been made by several groups that have targeted young objects in their RV searches of planetary companions. For example, surveys are being carried out which focus on both nearby associations of young stars and
moving groups with ages ranging 10--500 Myr; examples of which include $\beta$ Pic (12 Myr), UMa
association (300 Myr), Pleiades (100 Myr), IC 2391 (35 Myr), Hyades (700
Myr), Taurus association (2 Myr), ChaI (2 Myr), TWA (10 Myr) (Paulson et al. 2004, Paulson \& Yelda 2007, Esposito et al. 2006, Huerta
et. al 2007, Setiawan et al. 2007, Setiawan et al. 2008, Prato et
al. 2008).  Positive identification of planetary signatures from these
efforts are few, with only two candidates to date:
HD 70573 (Setiawan et al. 2007) and the controversial TW Hya (Setiawan
et al. 2008).
Planets orbiting around young stars are particularly valuable as they
enable us to investigate some of the critical questions about the
formation of both stellar and planetary systems. How, and at what
stage planets form, what is the planet formation mechanism, and how
they evolve are important questions which the study of young planetary
systems will help to answer.
In this paper we report strong evidence of a planetary candidate
orbiting the young and active K5V star BD+20 1790. Sect.~2, is an
overview of the properties and our previous studies of this star. The
observational strategy and data analysis are presented in Sect. ~3. In
Sect.~4 the nature of RV variations is investigated.  An orbital
solution for the data is presented in Sect.~5, and in Sect.~6 a
discussion about planetary parameters, orbital solution, and how
stellar activity and the planet are related is shown. Finally, we
summarize and offer some concluding remarks in Sect.~7.

\section{BD+20 1790: An overview}

BD+20 1790 was classified by Jeffries (1995)
as a K5Ve star, with a magnitude of $V = 9.9$. Mason et al. (1995)
identified this star as the optical counterpart for the 2RE
J072343.6+202500 EUV source, located in the ROSAT All-Sky Survey.
L\'opez-Santiago et al. (2006)
proposed its membership in the AB Dor kinematic moving group which has
an estimated average age of 50 Myr. By comparing the equivalent width
of Li $\lambda$ 6708 \AA \ with the spectral type, L\'opez-Santiago et
al. (2006) derived an age estimate of 35--80 Myr.
The main stellar parameters for BD+20 1790 are compiled in
Table~\ref{star}. We obtained a value for the stellar radius from the
measured rotational velocity and photometric period.
Our estimated radius agrees with the previous K5V spectral
classification (from Carrol \& Ostlie (2007) tables). Adopting this
spectral type, we used the K5V temperature from the Carrol \& Ostlie
(2007) tables. In conjunction with the photometric parameters, this
enabled us to derive the luminosity, mass and surface gravity. Errors in the parameters were estimated by following the method of
propagation of errors, i.e., the uncertainties were calculated from the
errors in the variables involved in the determination of each parameter.
It has been assumed null correlation between the different variables, in
principle independent of each other.
In order to test if assuming a fixed value for $T_{\rm eff}$ has a
non-negligible effect
in the errors computation, we investigated whether an error in $T_{\rm eff}$
could translate into
uncertainties of derived parameters.
We have considered an input error in $T_{\rm eff}$$\sim$10 $K$ and conducted an analysis
in the propagation of $T_{\rm eff}$ error. Based on this analysis, we have noticed
that not consider the error in the $T_{\rm eff}$ leads to an underestimation in the
errors in mass and $\log g$, providing unreliable error bars for
these parameters.

The X-ray luminosity was calculated using the count rates and HR1
hardness ratios from the ROSAT All-Sky Survey. By combining the
conversion factor $C_x$, computed by the formula from Fleming (1995),
and the distance estimated by Reid et al. (2004), the stellar X-ray
luminosity was calculated as $L_X$ = 1.6$\pm$ 0.5 $10^{29}$ $erg$ $s^{-1}$.\\
We compute a preliminary value of metallicity by using a grid of
Kurucz et al. (1993) ATLAS9 atmospheres and the 2002 version of
MOOG\footnote{The source code of MOOG 2002 can be downloaded at
  http://verdi.as.utexas.edu/moog.html} synthesis code (Sneden 1973).
Atmospheric models were constructed with the data given in
Table~\ref{star}.  We used 12 Fe I lines selected from Gonz\'alez et
al. (2001). We also calculated a value of metallicity by using 7 Fe I
lines in the MOOG Abfind routine. We find an average value of
$A[Fe]$ = 7.82$\pm$0.20 which, when assuming a solar value of
$A[Fe]$ = 7.52, results in a $[Fe/H]$ = 0.30$\pm$0.20. 
As mentioned, this is a preliminary value, although compared with the average metallicity of stars of solar neighbourhood, we still could consider the star as metal-rich within the error bars.

In a recent paper Carpenter et al. (2008) derived the temperature, gravity and metallicity for BD+20 1790, being their values of $T_{\rm eff}$ = 4408 $K$ and $\log g$ = 4.50, very close to the corresponding values presented in Table~\ref{star}. We also pointed out that the difference between metallicity values may be explained by the fact that Carpenter et al. (2008) only assumed a fixed metallicity of $[Fe/H]$ = 0.0, but not actually compute it.

The Li I abundances analysis was also done in standard local
thermodynamic equilibrium (ETL) using MOOG and ATLAS9 in the same way
as with the metallicities.  Abundances were derived by fitting
synthetic spectra to the data.  To determine Li abundances we perform
a spectral synthesis around the Li I~6707~{\AA} resonance doublet,
fitting all spectra between 6702 and 6712 \AA, taking into account the
relation between Li6 and Li7 isotopes. We determine an average value
of lithium abundance of $log N(Li)$ = 1.03$\pm$0.04 (where $log
N(Li)$ = log (Li/H)+12).

\begin{figure}
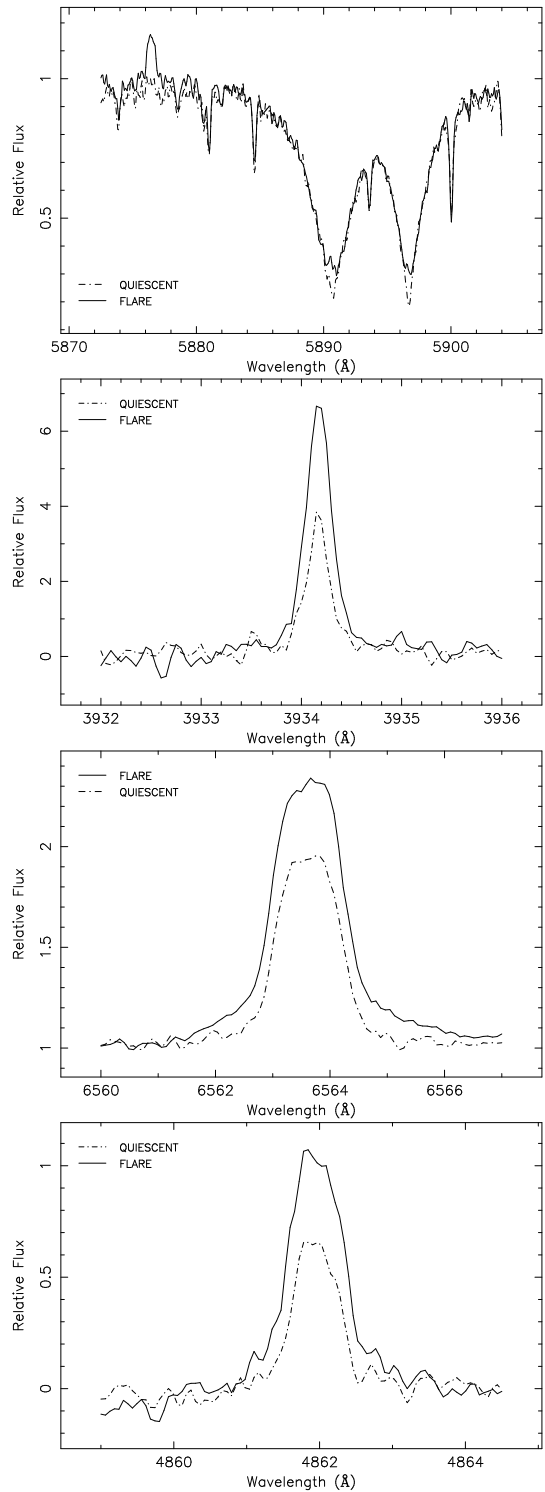

\includegraphics[angle=-90,scale=0.30,clip]{11000fg1a.ps}
\includegraphics[angle=-90,scale=0.30]{11000fg1b.ps}
\includegraphics[angle=-90,scale=0.30]{11000fg1c.ps}
\includegraphics[angle=-90,scale=0.30]{11000fg1d.ps}

\caption{Chromospheric activity indicators. The dashed line indicates quiescent state, while solid line indicates flare state. From top to bottom and left to right: He I D$_{3}$ region, Ca~{\sc ii} K, H$\alpha$ and H$\beta$}
  \label{fig:fig1}
\end{figure}

In order to study the stellar activity and the kinematics, we have
carried out both spectroscopic and photometric monitoring over the
past few years: high temporal and spectroscopic resolution and two
band photometry. The simultaneous study of photospheric and
chromospheric active regions is a powerful tool that allow us to
trace, reconstruct and model the puzzle of the magnetic field
topology, since these active regions are the fingerprints of magnetic
fields (Collier Cameron 2001, Catalano et al. 2002, Frasca et al. 2005,
Collier Cameron et al. 2002).
Strong chromospheric activity was detected in several observing runs,
described by Hern\'an-Obispo et al. (2005, 2007).
In spite of the fact that the rotational velocity is not very high, $v
{\rm sin}i\sim$10 km s$^{-1}$ (L\'opez-Santiago et al. 2006), all
activity indicators are in emission above continuum, from Ca~{\sc ii}
H \& K, to Ca~{\sc ii} IRT lines (see Fig.~\ref{fig:fig1}).\\ Through
the study of profile line asymmetries of H$\alpha$ and H$\beta$ lines,
prominence-like structures have been detected in the chromosphere of
the star (Hern\'an-Obispo 2005, 2007). These can be observationally
detected as transient absorption features superimposed on the line
profile that are interpreted as the presence of cool material embedded
in the surrounding hotter corona and co-rotating with star (Collier
Cameron \& Robinson 1989a,b, Collier Cameron \& Woods 1992, Jeffries
et al. 1993, Byrne et al. 1996, Eibe et al. 1998, Barnes et al. 2000,
Donati et al. 2001). Several completed prominence-like transients
have been detected with durations of orders of a few hours (see
Hern\'an-Obispo 2005 for details).
Modeling these chromospheric phenomenae is an important challenge in
this case, due to the detection of these prominence-like structures
in unstable positions, far from equatorial regions (Ferreira 2000,
Jardine et al. 2001, Jardine \& van Balegooijen 2006).

In addition, strong large optical flare events were observed. The
gradual decay of the flares was observed for up to 5
hours. Fig.~\ref{fig:fig1} compares the activity indicators for the
quiescent state and flare state. The energy released is on the order
of $\sim$10$^{37}$ erg, while for largest solar flares the released
energy is about $\sim$10$^{29}$--10$^{32}$ erg, thus ranging the
flares of BD+20 1790 on the so-called \textit {superflare} regime (Rubenstein \&
Schaefer 2000).

The photometric observations yielded a light curve with evidence of
rotational modulation, the semi-amplitude of which is up to
$\Delta${\it{V}}$\sim$ 0.$^m$06 and indicates the presence of spots on
the surface. The period analysis of the entire set of observations
reveals a photometric period of 2.801 ($\pm$ 0.001) days, in agreement
with the period given by the SuperWASP photometric survey (Norton et
al. 2007).

\begin{table}
\caption{Stellar Parameters of BD+20 1790}             % title of Table
\label{star}      % is used to refer this table in the text
\centering                          % used for centering table
\begin{tabular}{l r }        % centered columns (4 columns)
\hline\hline                 % inserts double horizontal lines
Parameter & Value   \\    % table heading
\hline                        % inserts single horizontal line
            Spectral Type & K5 V\\
            $B-V$ & 1.15 \\
            $M^{\mathrm{a}}$ & 0.63 $\pm$ 0.09 $M_{\sun}$\\
            $T_{\rm eff}^{\mathrm{b}}$ & 4410  K\\
            $\log g^{\mathrm{a}}$ & 4.53 $\pm$ 0.17 \\
            $EW{\rm(Li)}^{\mathrm{a}}$ & 110 $\pm$ 3  m\AA\\
            $Distance^{\mathrm{e}}$ & 25.4 $\pm$ 4 $pc$\\
            $Age^{\mathrm{c}}$ & 35 - 80  Myr\\
            $v {\rm sin}i^{\mathrm{d}}$ & 10.03 $\pm$ 0.47  km s$^{-1}$\\
            $P_{\rm phot}^{\mathrm{a}}$ & 2.801 $\pm$ 0.001  days\\
            $i^{\mathrm{a}}$ & 50.41  degrees\\
            $R^{\mathrm{a}}$ & 0.71 $\pm$ 0.03 $R_{\sun}$ \\
            $[Fe/H]^{\mathrm{a}}$ & 0.30 $\pm$ 0.20\\
            $log N(Li)^{\mathrm{a}}$ & 1.03 $\pm$ 0.04 \\
            $L_X^{\mathrm{a}}$ & 1.6 $\pm$ 0.5 $10^{29}$ erg s$^{-1}$ \\
            $L^{\mathrm{a}}$ & 0.17 $\pm$ 0.04 $L_{\sun}$ erg s$^{-1}$\\
\hline                                   %inserts single line
\end{tabular}
\begin{list}{}{}
\item[$^{\mathrm{a}}$] This paper
\item[$^{\mathrm{b}}$] From Carrol \& Ostlie, 2007
\item[$^{\mathrm{c}}$] From L\'opez Santiago et al. 2006
\item[$^{\mathrm{d}}$] From L\'opez Santiago 2005
\item[$^{\mathrm{e}}$] From Reid et al. 2004
\end{list}
   \end{table}

A detailed and completed study of the chromospheric and photospheric
activity characterization will be published in a forthcoming paper
(Hern\'an-Obispo et al. 2009b, in prep.).
%
%______________________________________________________________
\section{Observations and Data Analysis}

In order to study and characterize active regions at photospheric and
chromospheric levels, we carried out photometric and spectroscopic
observations of the target.

\subsection{Spectroscopic data}

The observational strategy was designed to spectroscopically monitor
chromospheric activity indicators with high temporal and spectral
resolution. High resolution echelle spectra were obtained during four
observing runs, from 2004 to 2007, detailed in Table~\ref{runs}. 
The exposure times ranged from 900 s to 1200 s, depending on weather
conditions, in order to obtain a S/N typically greater than 140 for
SARG runs and 80 for FOCES runs.
The spectra in the time series observations were separated only by the
CCD readout time, thus enabling us to obtain the highest temporal
resolution possible. Our initial temporal cadence was designed to
detect prominence-like transient features in the Balmer lines.
Spectral types and RV standards were acquired with the same setup and
configuration as the target. These standards were reduced and
analysed in the same way as the target. The data were bias-subtracted,
overscan-corrected and flat-fielded using standard routines in
IRAF\footnote{IRAF is distributed by the National Optical Observatory,
  which is operated by the Association of Universities for Research in
  Astronomy, Inc., under contract with the National Science
  Foundation.} package.

The wavelength calibration was obtained by taking spectra of a Th-Ar
lamp. Using Coud\'e spectrographs allowed a stable environment for
the wavelength calibration, since flexures are not possible. Details about the spectrographs
used can be seen in Pfeiffer et al. (1998) for FOCES spectrograph and Gratton et al. (2001) for SARG spectrograph.
In order to enhance the accuracy in calibration we used about 10--12
lines identified per order; across all orders for SARG spectra and
about 80 orders for FOCES spectra. The orders were calibrated
simultaneously and the total fit has an rms value typically lower than
0.003 \AA.
The spectra were normalized by a polynomial fit to the observed
continuum.

Heliocentric radial velocities were determined using a weighted
cross-correlation method. The spectra of the star were correlated
order by order against spectra of several RV standards with similar
spectral type. Orders with chromospheric features and telluric lines
were excluded. We calculated the uncertainties based on the
cross-correlation peak height and the antisymmetric noise as
described by Tonry \& Davis (1979).  Also, by measuring RVs of the
standard stars, we estimated the systematic errors and the accuracy of
the RV measurements with our instrumental setup. The accuracy between
standards for the same run and between runs is less than 0.05 km
s$^{-1}$.\\
Additional echelle data were acquired in DDT mode at the FOCES
spectrograph on December 2008. The telescope configuration and the
setup were identically to previous FOCES runs, except for two nights
in which a different CCD was used. Data were taken over 10 consecutive
nights but due to bad weather conditions only five nights were
acquired. Because of the time limitation in DDT mode, only one RV
standard was observed.

\begin{table*}
\caption[]{Observing runs
\label{runs}}
\begin{center}
\small
\begin{tabular}{lllllllll}
\noalign{\smallskip}
\hline  \hline
\noalign{\smallskip}
%Cabecera de la tabla
 Date & Telescope & Instrument & CCD chip & Spect. range & Orders &
Dispersion & FWHM$^{\rm c}$ & N. Obs.  \\
   &           &            &   \#     & ~~~~~~~(\AA)   &          & ~~~~~(\AA/pix) & ~~~(\AA) &  \\
\noalign{\smallskip}
\hline
\noalign{\smallskip}
%%.......................................................................
29/03-6/04 2004 & 2.2m$^{\rm a}$  & FOCES & 2048x2048 24$\mu$m Site$\#$1d & 3720 - 10850 & 100 & 0.04 - 0.13 & 0.08 - 0.35 & 19 \\
21-22/11/2004 & TNG $^{\rm b}$ & SARG & 2048x4096 13.5$\mu$m EEV & 4620 - 7920 & 52 & 0.07 - 0.11 & 0.07 - 0.17 & 43 \\
15/04/2006 & TNG $^{\rm b}$  & SARG & 2048x4096 13.5$\mu$m EEV & 4620 - 7920 & 52 & 0.07 - 0.11 & 0.07 - 0.17 & 14 \\
2-5/10/2007 & 2.2m$^{\rm a}$  & FOCES & 2048x2048 24$\mu$m Site$\#$1d  & 3720 - 10850 & 100 & 0.04 - 0.13 & 0.08 - 0.35 & 10 \\
12-13/12/2008 & 2.2m$^{\rm a}$  & FOCES & 2048x2048 15$\mu$m LORAL$\#$11i & 3830 - 10850 & 96 & 0.03 - 0.07 & 0.09 - 0.26 & 2 \\
19-21/12/2008 & 2.2m$^{\rm a}$  & FOCES & 2048x2048 24$\mu$m Site$\#$1d & 3620 - 7360 & 100 & 0.04 - 0.13 & 0.08 - 0.35 & 3 \\
%.......................................................................
\noalign{\smallskip}
\hline
\noalign{\smallskip}
\end{tabular}
\end{center}
\vspace{-0.25cm}
{\small
$^{\rm a}$ 2.2~m telescope at the German Spanish Astronomical Observatory (CAHA)
(Almer\'{\i}a, Spain).\\
$^{\rm b}$ 3.58~m {\it Telescopio
  Nazionale Galileo} (TNG) at Observatorio del Roque de los Muchachos
(La Palma, Spain).\\
$^{\rm c}$ The spectral resolution is determined as the FWHM at the arc comparison lines ranges. \\
}
\end{table*}

\subsection{Photometric data}

The purposes of these observations were to determine the photometric
period and to look for photometric variability. In addition, the study
of the light curve, as well as spectrocopy, allow us to characterize
the active regions in the photosphere (Catalano et al. 2002, Frasca et
al. 2005, Biazzo et al. 2007).

\begin{figure}
\centering
\includegraphics[angle=0,scale=0.92,clip]{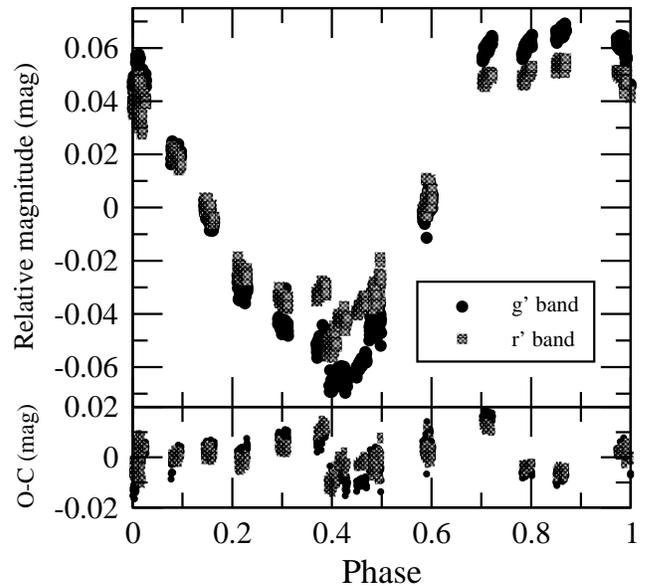}
        \caption{Photometry phased to the 2.801 days period. A linear trend and a zero point have
been subtracted to both bands The residuals with respect to a simple sine-wave
model are show in the lower panel.}
        \label{fig:fig2}
\end{figure}

\begin{figure}
\centering
\includegraphics[angle=0,scale=0.69,clip]{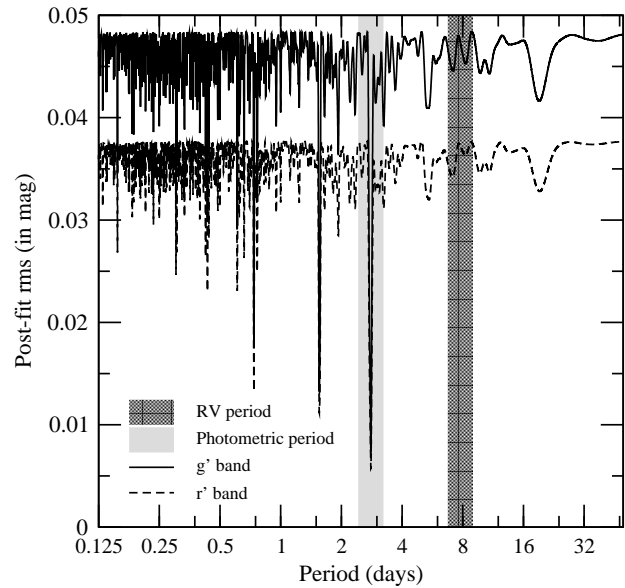}
        \caption{Postfit residuals to the photometry as a function of the period. The sharper
minima correspond to the 2.801 days period in both bands. The RVs period is
marked in grey to illustrate the absence of related photometric signals.}
        \label{fig:fig3}
\end{figure}

CCD differential aperture photometry was obtained using the $2.0$ m
fully robotic Liverpool Telescope (Steele et al. 2004) at the
Observatorio del Roque de los muchachos in La Palma, Spain. The
observations were scheduled in monitoring mode. We obtained 22
photometric epochs during November and December 2007. Our
observational strategy permitted us to obtain a photometric epoch
every 3 nights on average. Each epoch consisted in alternating
r$^\prime$ and g$^\prime$ exposures\footnote{Sloan r' and g' filters
  were used}, thereby obtaining quasi-simultaneous two band
photometry. Custom made software\footnote{ATP, Automatic TATOOINE
  Photometry. http://www.am.ub.es/$\sim$anglada/atp/atp\_testing.htm}
was used to automatically extract the photometry. By analysing intra
night scatter we can infer a photometric accuracy of 3 mmag and 4 mmag
per exposure ($r^\prime$ and $g^\prime$ bands respectively,
see Fig.~\ref{fig:fig2}).
We fit the best sine-wave model to the photometry sampling many
periods between 0.1 and 50 days, on both bands.
Plotting the post-fit residuals as a function of the period, a very
strong minimum on the post-fit residuals is found at $2.801 \pm 0.001$
days in both bands (see Fig.~\ref{fig:fig3}).  We note that the period
and the amplitude are similar with those given by the SuperWASP survey
(Norton et al. 2007).

The different amplitude in each band is consistent with large spot or
spot group covering at least $4\%$ of the surface. As can be seen in Fig.~\ref{fig:fig2}, the amplitude is larger at shorter wavelength, i. e. at $g^\prime$ band in this case. This color variation is correlated with variation in magnitude. The star appears redder when fainter, at minimum light and therefore bluer when brighter, at maximum light.\\
 The full analysis
of the photometry and its relation to the star activity requires
simultaneous discussion with the spectroscopic data and a more
detailed study of the star will be presented elsewhere
(Hern\'an--Obispo et al. 2009b, in prep.)

\section{On the nature of the RV variations}

Variations in the RV peak-to-peak amplitude of up to $\sim$2 km
s$^{-1}$ were observed during all the observing runs. These variations
are significantly larger than the individual measurement errors (0.10
to 0.20 km s$^{-1}$) or the systematic error (0.05 km s$^{-1}$), even
when we consider the scatter between runs with different spectrographs and
setups.

\subsection{Searching for periodical signals on RV}

\begin{figure}
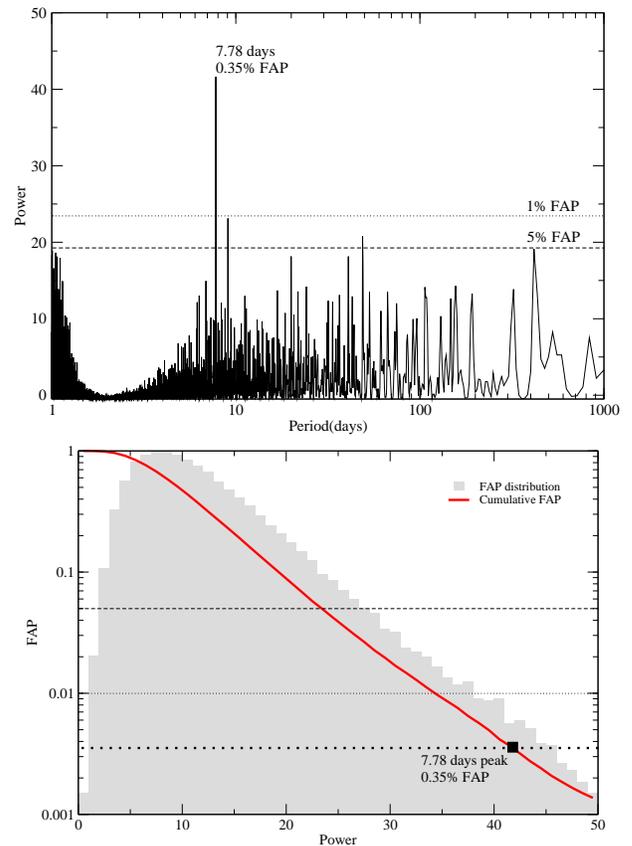

\centering
\includegraphics[angle=0,scale=0.34,viewport=0 0 675 480,clip]{11000fg4a.eps}
\includegraphics[angle=0,scale=0.32,viewport=0 0 690 480,clip]{11000fg4b.eps}
        \caption{\textbf {a. Up:} Least-Squares Periodogram of the nightly averaged radial velocity measurements. The 7.78 days peak has a FAP of $0.35\%$. The dotted horizontal line illustrates a FAP lower than $1\%$ and the dashed horizontal line a FAP lower than $5\%$.\textbf {b. Down:} Empirical False Alarm Probability as a function of the Power (red line). The gray bars illustrate the distribution of False Alarms with an arbitrary normalization used to derive the Empirical FAP. Note that the Y axis is in
logarithmic scale.
}
        \label{fig:fig4}
\end{figure}

A Least squares periodogram (see Appendix A) reveals one very
significant peak at 7.783 days (see Fig.~\ref{fig:fig4}a). The
data set contains 91 independent RV measurements. However, many of
them are clustered together within groups of a few hours. The values
we use to generate the periodogram and for orbital fitting (shown in
Table~\ref{rv}), are averaged on a nightly basis.
Fig.~\ref{fig:fig4}b shows the empirical False Alarm Probability (FAP) as a
function of the Power. The 7.783 days peak has a FAP of $0.35\%$.

It is worth noting that the RV period is larger than the photometric
period. Nevertheless to test if the RV period could arise from
rotational modulation we searched for significant frequencies in the
data points of the photometry. There is no significant power at the RV
period, and no secondary peaks are found in the aliasing frequencies
of the RVs or the photometric period after the main signals are
removed. To illustrate the absence of related photometric signals, we
marked the RV period in Fig.~\ref{fig:fig3}, that shows the post-fit
residuals of photometric data. In addition to this, there is no signal at photometric period in the RV data, as can be seen in Fig.~\ref{fig:fig4}a, that shows the RV periodogram.  

\begin{table}
\caption{Radial Velocity}             % title of Table
\label{rv}      % is used to refer this table in the text
\centering                          % used for centering table
\begin{tabular}{c c c}        % centered columns (4 columns)
\hline\hline                 % inserts double horizontal lines
JD days & RV (km/s) & $\sigma$ (km/s) \\    % table heading 
\hline                        % inserts single horizontal line
2452388.3341$^{\mathrm{a}}$ & 9.23 & 0.19 \\
2452389.3513$^{\mathrm{a}}$ & 8.94 & 0.14 \\
2452390.3670$^{\mathrm{a}}$ & 8.52 & 0.38 \\
2453099.3573$^{\mathrm{a}}$ & 7.82 & 0.06 \\
2453100.3692 & 6.96 & 0.10 \\
2453101.3748 & 7.34 & 0.07 \\
2453102.3876 & 7.84 & 0.05 \\
2454375.6480 & 7.96 & 0.08 \\
2454378.6804 & 7.72 & 0.04 \\
2453331.6400 & 8.71 & 0.03 \\
2453332.6800 & 8.16 & 0.03 \\
2453841.4250 & 7.73 & 0.03 \\
2454812.7429 & 7.76 & 0.21 \\
2454813.7240 & 7.67 & 0.18 \\
2454820.5057 & 7.73 & 0.16 \\
2454821.5126 & 7.53 & 0.16 \\
2454822.5483 & 7.96 & 0.14 \\
\hline                                   %inserts single line
\end{tabular}
\begin{list}{}{}
\item[$^{\mathrm{a}}$] From L\'opez-Santiago 2005
\end{list}
\end{table}

\subsection{Stellar activity jitter}

\begin{figure}
   \centering
   \includegraphics[angle=-90,scale=.34]{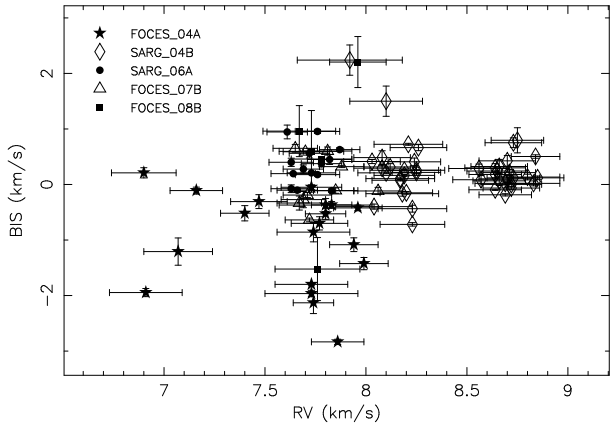}
      \caption{Bisector velocity span vs. Radial velocity for all the observing runs. Symbols represent the different runs: stars for FOCES 04A, diamonds for SARG 04B, circles for SARG 06A and
triangles for FOCES 07B. The lack of correlation indicates that RV variations are not due to
stellar activity.}
         \label{fig:fig5}
   \end{figure}

\begin{figure}
   \centering
   \includegraphics[angle=0,scale=.33]{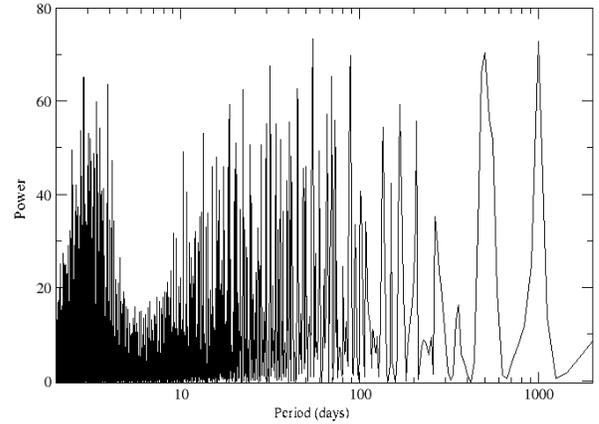}
      \caption{Periodogram for bisectors of all runs. There is no clear period for bisector variations.}
         \label{fig:fig6}
   \end{figure}

It is well known that spurious RV variations can be induced by stellar
activity, especially due to changes in the profile of spectral lines
caused by the presence of active regions, the so-called {\it stellar
  jitter} (Saar \& Donahue 1997, Saar 2009). The high level of
activity detected in BD+20 1790, induced us at first to relate RV
variations with active regions.
Since we ruled out the possibility of variations due to systematic
errors or any seasonal effect, the main concern was to determine if
stellar activity was responsible.\\ It is widely accepted that the
relationship of bisectors of the cross-correlation function (CCF) and
RV is a powerful method to determine whether the RV variation may be
due to stellar activity or a planetary companion (Queloz et al. 2001,
 Mart\'{\i}nez-Fiorenzano et al. 2005).
The CCF was determined by using the same procedure as for the RV case,
computing it for the regions which include the photospheric lines
which are more sensitive to spot presence, while excluding
chromospheric lines and telluric lines. The bisector inverse slope
(BIS), defined as the difference of the average values of the top and
the bottom zones, was computed to quantify the changes in the CCF
bisector shape by using the method described by Queloz et al. (2001). In
choosing the span zones we avoided wings and cores of the CCF
profiles, where errors of bisectors measurements are large. In
Fig.~\ref{fig:fig5} it can be seen that there is a lack of correlation
between the BIS and RV variation for all the observing runs. This
indicates that the RV variations are not due to variations in the
asymmetry of the photospheric lines profile, and subsequently not due
to stellar activity variations. The least squares periodogram of
bisectors shows two tentative peaks around 2.8 days, and is shown in Fig.~\ref{fig:fig6}.

\begin{figure}
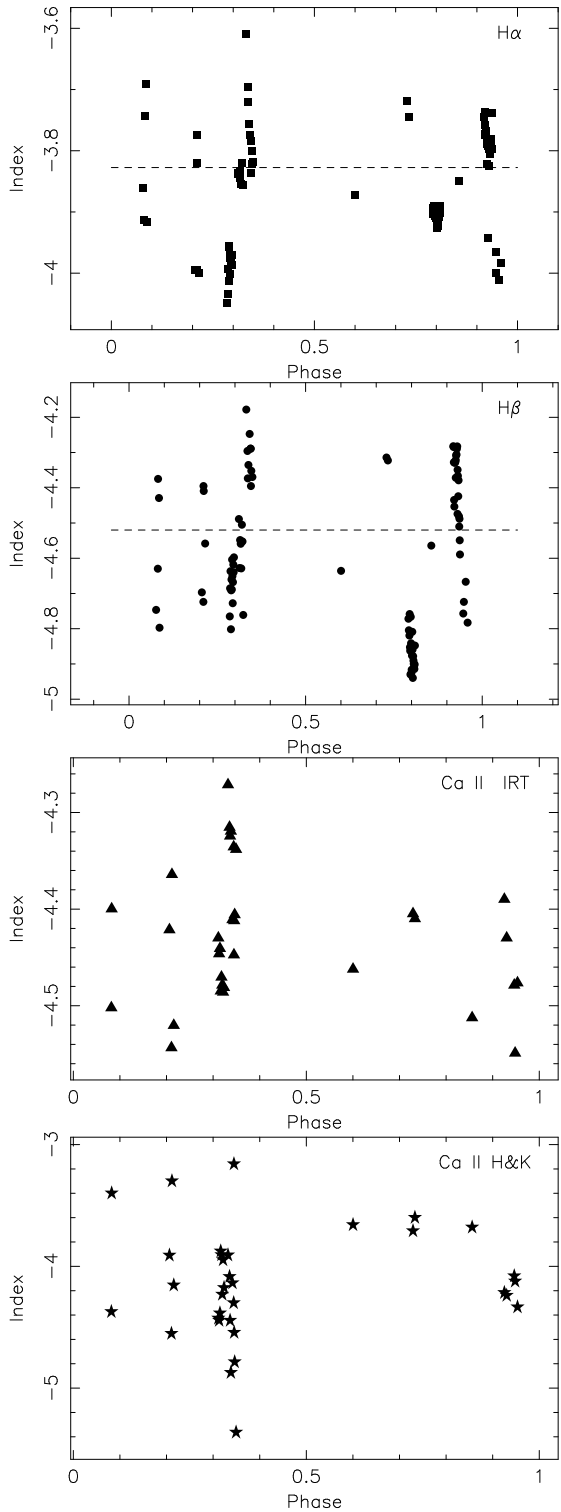

    \includegraphics[angle=-90,scale=0.32]{11000fg7a.ps}
    \includegraphics[angle=-90,scale=0.32]{11000fg7b.ps}
    \includegraphics[angle=-90,scale=0.32]{11000fg7c.ps}
    \includegraphics[angle=-90,scale=0.32]{11000fg7d.ps}
        \caption{Spectroscopic index for chromospheric activity indicators, phased folded orbital period. From top to bottom: H$\alpha$ (squares), H$\beta$ (circles), Ca~{\sc ii} IRT (triangles) and Ca~{\sc ii} H \& K (stars). The dashed line is indicating the quiescent state. Error bars for indices are of order of 0.001}
        \label{fig:fig7}
\end{figure}

\begin{figure}
\centering
\includegraphics[angle=0,scale=0.35,viewport=0 0 681 477,clip]{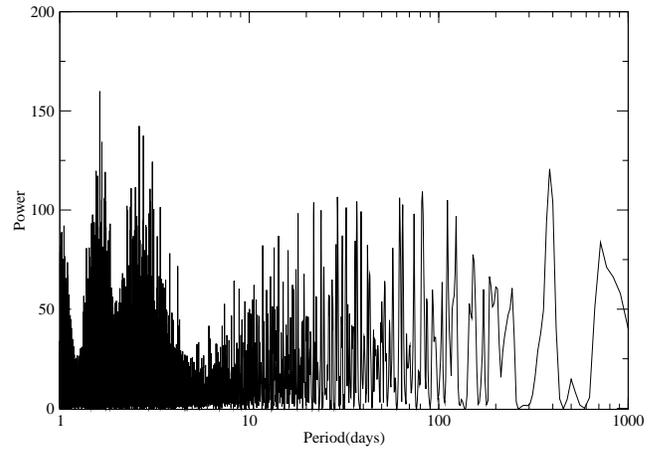}
        \caption{Periodogram for H$\alpha$ index. There is no clear period for index variations.}
        \label{fig:fig8}
\end{figure}

We have estimated the stellar jitter from Santos et al. (2000), that takes into account the Ca~{\sc ii} H \& K index. Assuming an average value for Ca~{\sc ii} H \& K index of about -4.2 and by using eq. [4], we derived a value for the stellar jitter of up to 10 m s$^{-1}$. This stellar jitter is added in quadrature to the RV error.\\  
As an additional test we investigated the variation of stellar
activity indicators, especially those that are ascribed to the
presence of plage-like structures on the chromosphere, like Balmer
lines, Ca~{\sc ii} H \& K and Ca~{\sc ii} IRT. The emission flux for
these lines in active stars usually shows a periodic modulation (and
subsequently the spectroscopic indices) which is most likely due to
rotational modulation of plage-like structure emission. As is shown
in Sect.~2, all chromospheric activity indicators are in emission above
the continuum, indicating a very high level of activity. To avoid the
photospheric contribution to the spectral profiles we applied the
spectral subtraction technique described in detail by Montes et
al. (1995). This technique makes use of the program {\sc STARMOD}
developed at Penn State University (Barden 1985) and lately
modificated by Montes et al. (1995).  Also, in order to control the
error and minimize the uncertainties, some routines of the
astronomical data reduction package
\reduceme\ \footnote{http://www.ucm.es/info/Astrof/software/reduceme/reduceme.html}
developed at Universidad Complutense de Madrid (Cardiel 1999) were
used.  In these subtracted spectra, spectroscopic indices have been
defined and computed following Saar \& Fisher (2000), K\"uster et
al.(2003), Bonfils et al. (2007). Both, Ca~{\sc ii} IRT and Ca~{\sc
  ii} H \& K indices were only determined for FOCES runs, due to the
wavelength range coverage of the spectrograph. To avoid contamination
from telluric lines we only consider the 8662\AA\ Ca~{\sc ii} IRT
line.  We searched for periodic signals in the spectroscopic indices
by computing their Least squares periodograms. Fig.~\ref{fig:fig7}
shows the variation with time (orbital phase folded in this case)
for Ca~{\sc ii} IRT, Ca~{\sc ii} H \& K, H$\alpha$ and H$\beta$
indices. The corresponding periodogram computed shows more noise
rather than a clear signal. This result is also seen in the indices
figures as a non-modulation of the activity index. As an example,
Fig.~\ref{fig:fig8} shows the periodogram for the H$\alpha$ index.

\begin{figure}
\centering
\includegraphics[angle=-90,scale=0.36]{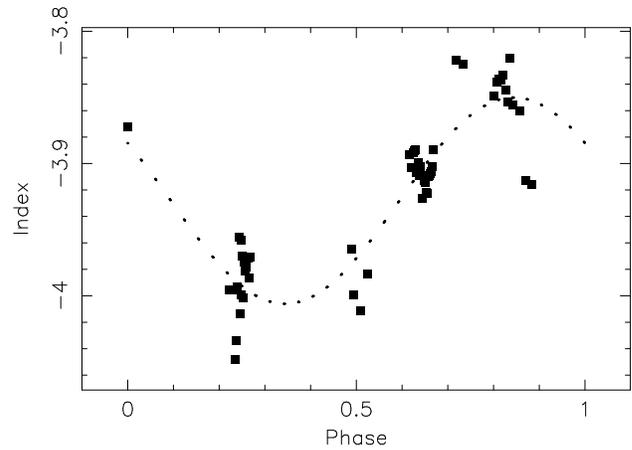}
        \caption{H$\alpha$ index for the data without flare events. It can be seen a modulation with a period of about 2.77 days, similar to photometric period.}
        \label{fig:fig9}
\end{figure}

\begin{figure}
\centering
\includegraphics[angle=-90,scale=0.34]{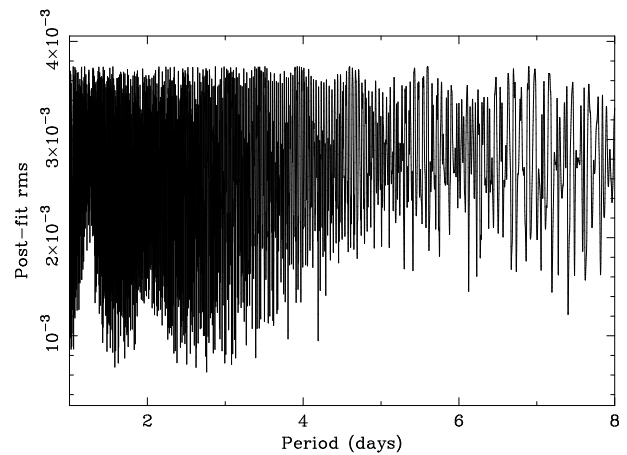}
        \caption{Postfit residuals to H$\alpha$ index for no-flare data as a function of the period. There is no clear period for index variations.}
        \label{fig:fig10}
\end{figure}

As pointed out by Walter (1994), the rotational modulation of
chromospheric lines due to plages is not always detectable in very
active stars. Furthermore, in this case the flares could contaminate
the data, masking the actual period of variation of the indices. 
In order to investigate this possibility we removed the data affected by flare events.
 Due to the different wavelength range coverage of spectrographs, we considered only H$\alpha$ and H$\beta$ indices. For H$\alpha$ index, we have found a tentative rotational modulation with a period of 2.77 days, similar to photometric period (see Fig.~\ref{fig:fig9}). However, the postfit residuals show in Fig.~\ref{fig:fig10} that this could be a misleading signal, even pure noise. For H$\beta$ index, no clear modulation has been found.\\
The lack of variability of BIS and spectroscopic indices with RV
period, and the absence of a photometric period larger than $2.8$
days, strongly support the planetary companion hypothesis.

\subsection{RV wavelength dependence}

Desort et al. (2007) (hereafter D07) pointed out that the color
dependence (with wavelength) of the RV peak-to-peak amplitude with
spots can be used as a diagnostic to discriminate between stellar
activity or planetary companions. Due to the contrast between spots
and the surrounding photosphere being greater in the visible than at
IR wavelengths, it is expected that an attenuation of RV amplitude
towards red wavelengths would be seen. Observationally this effect has
been shown by e.g. Mart\'in et al. (2006), Hu\'elamo et al. (2008) and Prato et al. (2008). If
the RV variations are due to a planet, the RV amplitude should be the
same at every wavelength range.  We investigated a possible chromatic
dependence by computing the RV in two different ranges of wavelength,
one for red and near-IR wavelengths (7650 to 10000 \AA) and the other
for blue (4300 to 4800 \AA).  The resulting RV peak-to-peak amplitude
is 2.19$\pm$0.20 km s$^{-1}$ for the near-IR range and 2.20$\pm$0.20
km s$^{-1}$ for the blue range.
The values differ by only 0.5\% and agree within the uncertainties.
Additional RV infrared follow-up can allow us to confirm this. In a
forthcoming paper (Hern\'an--Obispo et al. 2009d, in prep.), we will
present the first results of the study of the RVs of BD+20 1790 in the
near-IR range.
 
\subsection{RV variation by empirical spots and plages?}

To estimate an order of magnitude of the expected RV amplitude due to
spots, we use empirical relations derived by Saar \& Donahue 1997
(hereafter SD97) and D07. These relations connect the RV amplitude
with the spot filling factor {\rm $f_s$} and $v{\rm sin}$i.
We consider both relations by D07 and SD97, because D07 relations take
into account the spectral type and the whole spectral range (except
telluric and chromospheric lines) to compute the empirical RV, whereas
SD97 uses a single line and G5V spectral type.  Using SD97 eq. [1], we
derived an amplitude of up to 575 m s$^{-1}$ and by using D07 eq [5]
we similarly estimate an amplitude of up to 600 m s$^{-1}$. As mentioned, these results 
are taken as a quantitative estimation. There are more effects that are not taken into account here, like the spot location at stellar surface given by the colatitude $\theta$, and the spot temperature. SD97 eq. [1] and D07 eq. [5] considered the simple case of an equatorial spot, but SD97 assumed a $T_{spot}$ = 0 K and D07 assumed an spot temperature 1000 K cooler than the photosphere. The difference between the RV amplitude derived from both equations could be due to this different spot temperature.

On the other hand, we can estimate the spot filling factor that could produce
the RV signal of our data. We considered an average semi-amplitude of 1
km s$^{-1}$. The {\rm $f_s$} estimated from SD97 is therefore 23\% while D07
indicates 19\%. The {\rm $f_s$} measured from photometric variation is about 4\%.
 These results indicate that the spot filling factor needed to
explain the RV variation due purely to spots is not in agreement with
the photometry.

Saar (2003) and Saar (2009) showed significant efforts to model
plage-induced RV jitter. Although the models are mostly applicable to
solar-like stars, we could estimate the plage filling factor {\rm
  $f_p$} that could produce the RV signal by using the Saar (2009) equation
that connects the RV amplitude with $v{\rm sin}$i $>$ 6 km s$^{-1}$.
This {\rm $f_p$} estimated is about 70\%, that strongly suggets that the RV
variation is not due to chromospheric plages.

\subsection{What would the RV signal be without a planet?}

It is important to remark that empirical relations derived by SD07 and
D07 do not take into account the chromatic effect of spots on the RV
signal.
We therefore investigated how much RV signal
would be expected in the absence of a planet, and the degree of
RV attenuation with wavelength (assuming the RVs are due to cool spots). In order to quantify the attenuation
if the cause of variations were spots, we try to investigate how much
spots affect the line profiles.  However, BD+20 1790 has a low $v {\rm
  sin}$i to model the photosphere by generating Doppler imaging spot
maps.  To carry out a realistic approximation to the problem, we
construct realistic spot maps by using spectra of another star with
similar characteristics, LO Peg, that is widely studied in the literature
and its photospheric activity is well-known (Jeffries \& Jewell 1993,
Jeffries et al. 1994, Eibe et al. 1998, Eibe et al. 1999, Barnes et
al. 2005).  LO Peg is a K5V--K7Ve star, identified by Jeffries \& Jewell (1993) as
a member of the Local Association, with an estimated age of 20--30
Myr. Jeffries et al. (1994) determined the inclination to be
$50^{\circ}$. The level of activity is similar to BD+20 1790, but LO
Peg is a rapid rotator ($v {\rm sin}i$ $\sim$69 km s$^{-1}$). The LO
 Peg photometry suggests a spot filling factor of up to 1.5\%.

Using the Doppler imaging program, DoTS (Collier Cameron 1997), and an
input starspot image derived for LO Peg (Barnes et al. 2005), we
generated a set of line profiles for a star with vsini = 10 km/s (i.e.
matching that of BD+20 1790) over a complete rotation phase. The
profiles thus contain asymmetries due to starspots from the observed
LO Peg image. We used appropriate temperatures for the BD+20 1790
photosphere and estimated the spots to possess temperatures which were
up to 1000 K cooler.  Profiles were generated for the three different
wavelengths of 4000 \AA, 6717 \AA~and 10000 \AA. The radial velocity
variations were then calculated in order to estimate the relative
amplitudes due to spot induced variations at each of the three
wavelengths.

\begin{figure}
\centering
\includegraphics[angle=-90,scale=.33]{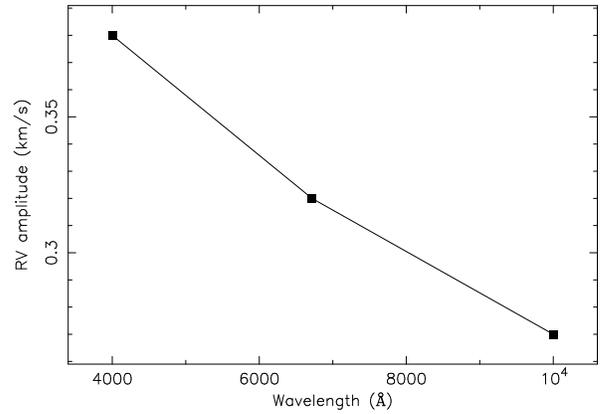}
     \caption{Radial velocity amplitude variation with wavelength, computed for LO Peg profiles}
        \label{fig:fig11}
   \end{figure}

The RV attenuation with wavelengths relative to 4000 \AA~is 16\% at
6717 \AA~and about 30\% at 10000 \AA~, as illustrated in Fig.~\ref{fig:fig11}. Assuming 
at a first approach the same {\rm $f_s$} for LO Peg and BD+20 1790, the RV signal for
BD+20 1790 should be about 1.5 km s$^{-1}$ at 10000 \AA.

However, in Hern\'an--Obispo et al. 2009d (in prep.), we find only
about 0.5\% attenuation in the near-IR region relative to the visible,
6717 \AA~region.  This result is an additional argument in support the
existence of the planetary companion.

\subsection{RV jitter from flares}

We have estimated the rate of flare occurrence as the fractional
amount of the total observing time (for all runs) where a flare was
detected. Thus, we get a flare frequency of occurrence of $\sim$
40\%. This higher rate raises the question of how much RV jitter we
should expect from large flares, if any. Saar (2009) presents the first
approach to this issue, concluding that RV jitter due to flare
occurrence would be non-negligible, although probably be a stochastic
jitter component.  Chromospheric activity indicators exhibited an
enhancement at flare state, the broad emission of Balmer lines and He
I $D_{3}$ in emission being the most notable features (see
Fig.~\ref{fig:fig1}). As pointed out by Saar (2009), although these
lines are excluded when we measure RV, it is possible that a
significant core filling in photospheric lines occurs when there is a
flare event. The cause could be upper photospheric heating. Results by
Houdebine (1992) state that heating is propagated down to low
photospheric levels.

\begin{figure}
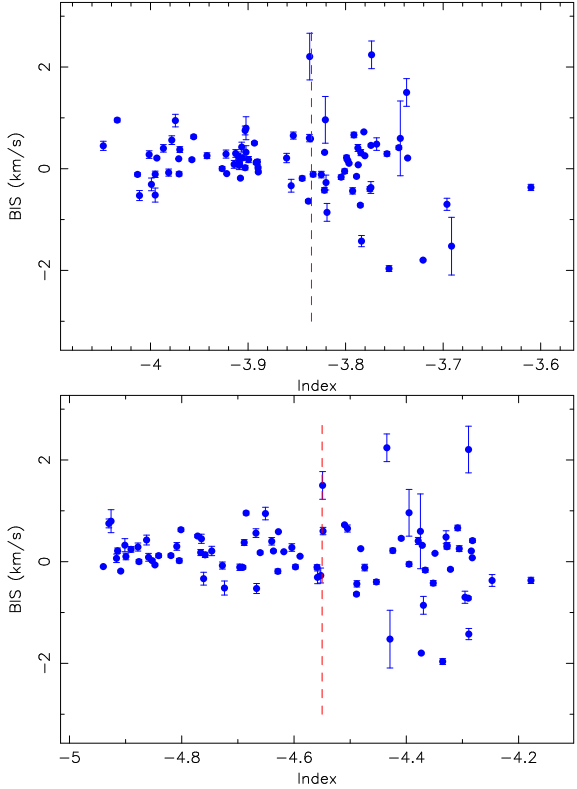

   \centering
   \includegraphics[angle=-90,scale=.32]{11000fg12a.ps}
   \includegraphics[angle=-90,scale=.32]{11000fg12b.ps}
        \caption{\textbf {Up}: H$\alpha$ index vs. BIS. The dashed line is indicating the quiescent state.\textbf{ Down}: H$\beta$ index vs. BIS. For both, it is seen that the scatter for BIS is higher when occur flare events. Error bars for the indices are of order of 0.001.}
        \label{fig:fig12}
   \end{figure}

A second related problem is the effect of large flares on BIS. While
it has not been studied until now, it is expected to be more
pronounced, since bisectors are more sensitive to changes in line
profiles. To our knowledge it is reported here for the first
time. Fig.~\ref{fig:fig12}~a shows the relationship of H$\alpha$ index
vs. BIS, where the dashed line corresponds to the quiescent state, and
higher values for H$\alpha$ index indicate the occurrence of a flare
event. It is seen that the scatter for BIS is higher when a flare
occurs. Outliers at quiescent state correspond to a low S/N
rate. Similar BIS behaviour is seen in Fig.~\ref{fig:fig12}~b, that
shows H$\beta$ index vs. BIS.

\begin{figure}
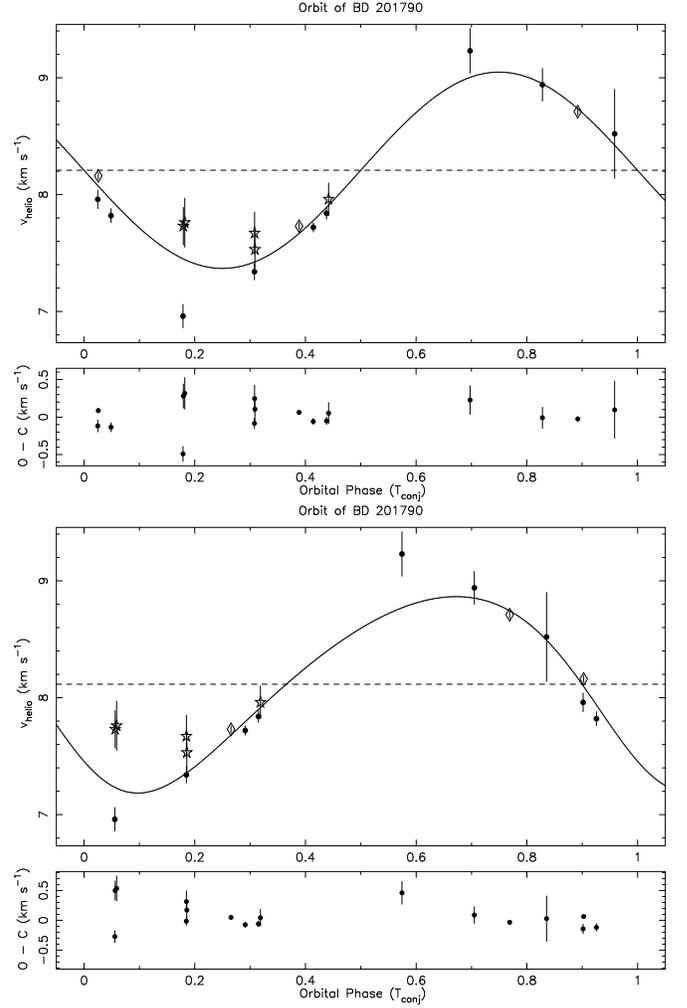

\centering
\includegraphics[angle=-90,scale=.345]{11000fg13a.ps}
\includegraphics[angle=-90,scale=.345]{11000fg13b.ps}
        \caption{Radial velocity variability of BD+20 1790.\textbf{a. Up}: Circular orbit. \textbf {b. Down}: Eccentric orbit. Values marked with circle symbol represent FOCES runs except stars that represent DDT FOCES 08B run. Diamond symbol are for SARG runs.}
        \label{fig:fig13}
   \end{figure}

\section{Orbital solution for BD+20 1790 b}

We computed the orbital solution for the RV data using a standard
Keplerian fit with the RV period estimated by the Least squares
periodogram.  The fit was obtained firstly only considering the FOCES
data, averaged by night, in order to avoid intra-night scatter. After
this, we added the SARG data to improve the fit. The results for the fit considering only FOCES data or all data from the two 
spectrographs were compatible within uncertainties. With the addition of
RVs measured in winter 2008 (DDT FOCES 08b run), the Least squares
periodogram is strikingly improved, and the 7.78 day peak clearly
dominates the power spectrum.  Attempts to perform a Keplerian fit
using the second and the third highest periodogram peaks produced
significantly worse folded curves.  We have included to perform the fit the RV set computed by L\'opez--Santiago (2005). A first fit (see Fig.~\ref{fig:fig13}a) derives a close-in massive
planet ($a =0.066$~AU, $M_2 \sin i = 6.54 M_{jup}$) in a circular
orbit ($e = 0.05$) with a rotational period of 7.7834 days and a
reduced $\chi^2$ of 1.07. Also we present a second fit (see Fig.~\ref{fig:fig13}b) with the same
period for an eccentric orbit ($a = 0.066$~AU, $M_2 \sin i = 6.15
M_{jup}$, $e = 0.14$, $\chi^2 = 0.997$). Due to the sampling of the
data, we cannot discard a possible eccentric orbit.

Orbital elements for both solutions are compiled in Table~\ref{planet}
and discussed in the next section.

\begin{figure}
\centering
\includegraphics[angle=-90,scale=.345]{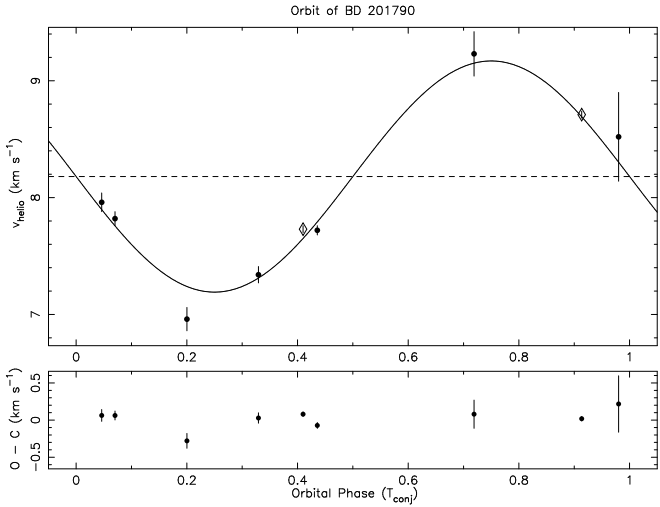}
        \caption{Radial velocity variation of BD+20 1790 computed considering only the data that are not affected by flares. Circle symbol represent FOCES runs except stars that represent DDT FOCES 08B run. Diamond symbol are for SARG runs.}
        \label{fig:fig14}
   \end{figure}

As additional test, we computed the orbital solution removing the data affected by flare events. The fit derives a solution ($a =0.066$~AU,  $M_2 \sin i = 6.54 M_{jup}$, $e = 0.01$, $K = 0.91$ km s$^{-1}$) compatible with the solution when considering all the data. The fit is presented in Fig.~\ref{fig:fig14}.

\begin{table}
\caption{Orbital Parameters of BD+20 1790 b}             % title of Table
\label{planet}      % is used to refer this table in the text
\centering                          % used for centering table
\begin{tabular}{l c c c}        % centered columns (4 columns)
\hline\hline                 % inserts double horizontal lines
Parameter & Solution 1 & Solution 2 &  \\    % table heading
\hline                        % inserts single horizontal line
$P_{\rm orb}$ & 7.7834 $\pm$ 0.0004 & 7.7834 $\pm$ 0.0004  & days \\
$T_{\rm conj}^{\mathrm{a}}$ & 3085.8 $\pm$ 0.5 & 3086.30 $\pm$ 0.18 & \small HJD  \\
$a$ & 0.066 $\pm$ 0.001 & 0.066 $\pm$ 0.002 &AU \\
$e$ & 0.05 $\pm$ 0.02 & 0.14 $\pm$ 0.04 &\\
$K$ & 0.93 $\pm$ 0.03 & 0.84 $\pm$ 0.06 & km s$^{-1}$\\
$\gamma$ & 8.22 $\pm$ 0.01 & 8.12 $\pm$ 0.04 & km s$^{-1}$\\
$\omega$ & 200.4 $\pm$ 21.8 & 120.7 $\pm$ 14.0 & \small degrees  \\
$M_2 sini$ & 6.54 $\pm$ 0.57 & 6.15 $\pm$ 0.59 & $M_{jup}$\\
$rms$ & 138.9 & 132.3  & m s$^{-1}$\\
$\chi^2$ & 1.071 & 0.997 & \\

\hline                                   %inserts single line
\end{tabular}
\begin{list}{}{}
\item[$^{\mathrm{a}}$] Time of periastron passage
\end{list}
\end{table}

\section{Discussion}

The lack of a relation between the BIS and spectroscopic indices with
the RV period, as well as the different RV and photometrical period
strongly suggest that the RV variations are due to a planetary
companion.  However, it is possible that the RV variations are
actually due to a combination of phenomena (activity and
planet).\\ Stellar magnetic activity may be influenced and enhanced by
the presence of a close-in giant planet, as proposed by Cuntz et
al. (2000), Cuntz \& Shkolnik (2002) and Lanza (2008). Thus, the presence of a planetary
companion could be an interpretation for the high level of stellar
activity detected. In a recent paper, Lanza (2009) proposes a new
model that predicts the formation of prominence-like structures in
very highly active stars with close-in giant planets. Also, as
presented in Sec.~2 and Sec.~4.6, the large flares, with energy
releases in the superflare regime, and the high rate of flare
ocurrence, could find a source in addition to stellar activity in the
reconnection of the stellar coronal field as the planet moves inside
the Alfv\'en radius of the star (Ip et al. 2004). In a forthcoming
paper we explore in detail these possible star-planet interactions
(Hern\'an--Obispo et al. 2009c, in prep).

In addition, as suggested by the statistical analysis by Kashyap et
al.(2008), the X-ray flux from stars with close-in giant planets is on
average 4 times greater than those with more distant planetary
companions. For the 'close-in' sub-sample, the X-ray luminosity is
$L_X$ = $10^{28.5}$ $erg$ $s^{-1}$ on average. The X-ray luminosity of
BD+20 1790 is 5 times brighter than this average which is consistent
with chromospheric and X-ray emission induced by the presence of a
massive close-in companion (Lanza 2009).\\ Even though the stellar
activity could swallow the RV signal of a planetary companion, we can
detect it for BD+20 1790 b since it is a massive planet. The RV
variation is large enough even though the RV accuracy is tipically
about 150 m s$^{-1}$.

Due to the observational strategy (the data are not part of a
planet-search program), the eccentricity is poorly constrained. Indeed
there is no "a priori" reason to discard an eccentric orbit since the
circularization time-scale computed is up to several Gyr, but more
data is required to properly characterize the eccentricity. RV optical
and infrared follow-up over twice the RV period will enable us to
constrain the orbital solution as well as confirm the presence of the
planet.  More massive exoplanets $M_2\sin i\sim$5$M_{jup}$ with
orbital periods longer than about 6 days have eccentricities
significantly larger than lower mass planets (Udry \& Santos 2007).
Another possibility is that additional undetected longer period
planets are maintaining the eccentricity of BD+20 1790 b. Both
situations have been discussed in detail by Wu \& Murray (2003).
\\ It is worth noting however that the star is metal-rich, as presented in Sec.~2. The
existence of a correlation between stellar metallicity and planet mass
has been reported by e.g. Santos et al. (2001), Fischer \& Valenti
(2005), Guillot et al. (2006). Massive planets tend to form around
metal-rich stars, i.e., planets that orbit around metal-rich stars
also have higher mass cores. \\
Compared to other planets of similar masses and
orbits\footnote{Observational data for the more than 370 exoplanets
  are compiled on the {\it Extrasolar Planets Encyclopaedia} ({\it
    http://exoplanet.eu}), mantained by J. Scheneider}, and taking
into account statistical results described in recent reviews (Udry \&
Santos 2007), BD+20 1790 b does not exhibit unusual characteristics,
except for its young age and its relatively high mass.  We used a
complimentary method to determine the stellar age from Mamajek \&
Hillenbrand (2008) (hereafter MH08), that uses the fractional X-ray
luminosity, $R_X$ = $L_X$/$L_{bol}$. MH08 demonstrate that $R_X$ has
the same age-inferring capability as the chromospheric index
$R^{\prime}_{HK}$. By using their equation [A3] we estimated an age
for BD+20 1790 of up to 35 Myr. Considering a value for $log
R^{\prime}_{HK}$ = -4.2 on average, we can also estimate the age with
the new relation proposed by MH08, by equation [3]. We computed an age
of up to 58 Myr. These values are in agreement with the range
estimated by L\'opez-Santiago et al. (2006).

Lowrance et al. (2005) included this star BD+20 1790 {in} a
coronographic survey for substellar companions using the coronograph
on NICMOS/HST and the 200 inch Hale Telescope (Palomar
Osbservatory). No companions were found beyond $10$ AU. However, the
orbital solutions we find suggest a semi-major axis below $0.1$ AU,
clearly beyond their resolving capabilities.

Great care must therefore be taken when
extrapolating properties of early stellar evolution stages from the
characteristics of the latter stages, since the current knowledge
about planetary system evolution is still somewhat speculative.
The exoplanetary zoo is such that new planets with unusual
properties require a replanting of planet formation and migration
scenarios. Planets discovered around young stars could be the missing
link that reconstruct the scenarios between exoplanets and
protoplanetary disks. Indeed, further study of BD+20 1790 b has the
potential to improve our understanding of planetary systems at early
evolutionary stages.

\section{Conclusions}

This paper describes the investigation of RV variations for the young
and active K5V star BD+20 1790. Based upon the analysis of the BIS of
the CCF, as well as activity indicators and photometry, the presence
of a planetary companion is shown to be the best interpretation.  The
orbital solution results in a companion with a mass in the planetary
regime. No photometric period larger than $2.8$ days strongly supports
the planetary origin of the observed RV variations. Two solutions for
the orbit are computed and discussed. The presence of a close-in
massive planet could also be an explanation for the high level of
stellar activity. Since the RV data are not part of a planet search
program, we can consider our results as serendipitous evidence of a
planetary companion. Indeed additional RV optical and infrared
follow-up will enable us to constrain the orbital solution as well as
confirm the presence of the planet. This is thus far the youngest main
sequence star for which a planetary candidate has been reported.

\begin{acknowledgements}
We thank Calar Alto Observatory for allocation of director's
discretionary time to this programme.  This work was supported by the
Spanish Ministerio de Educaci\'on y Ciencia (MEC) under grant
AYA2005-02750, Ministerio de Ciencia e Innovación (MICINN) under grant
AYA2008-06423-C03-03 and “The Comunidad de Madrid” under PRICIT
project S-0505/ESP-0237 (ASTROCAM). MCGO acknowledges financial support from the European
Commission in the form of a Marie Curie Intra European Fellowship
(PIEF-GA-2008-220679). MHO and GAE thank Dr. Chriss
Moss, support astronomer at the LT for his help and patience. Also MHO thanks Dr. Santos Pedraz, support astronomer at the Calar Alto Observatory for his help with DDT run. MHO is grateful to Dr. Jos\'e Antonio Caballero for valuable
discussions, and also Dr. Laurence R. Doyle for his suggestions that
was the initial inspiration for this work.  This research has made use
of the SIMBAD database, operated at CDS, Strasbourg, France.  The
authors gratefully acknowledge the valuable comments and suggestions
of an anonymous referee, that are helped to improve the paper.
\end{acknowledgements}

\begin{appendix}
\section{Periodogram}
We use a Least squares periodogram approach to identify and visually
illustrate the relevant periods in the data. It differs from the more
classic Lomb-Scargle periodogram (Scargle 1982) in a few key
aspects. For a given period $P$, a linear model of the form $v_r =
\gamma + A\cos 2\pi/P t + B\cos 2\pi/P t$ is fitted using a weighted
least squares to the data and the $\chi^2$ of the residuals is
obtained. The $\chi^2$ minima reveal the candidate signals of
interest. One can represent the root mean square of the residuals
(RMS) with respect to the period to show the relevant periods as
minima (This approach is used to illustrate the photometric periods in
Fig.~\ref{fig:fig3}). To recover a more familiar view of a
periodogram, one can compute the \textit{Power} of each period $P$ as

\begin{eqnarray}
{\rm Power}(P) = \frac{(\chi^2_{\rm
    none}-\chi^2_{P})/2}{\chi^2_{P}/(n_{\rm obs}-3)}\\ \chi^2_{\rm
  none} = \sum_{\rm i}^{\rm obs}
\left(\frac{v_i-\left<v\right>}{\sigma}\right)^2
\end{eqnarray}

that follows a Fisher-F Distribution with $2$ and $n_{obs}-3$ degrees
of freedom and can be used to obtain a first hint of the False Alarm
probability of a given solution. This definition of the power measures
how much the $\chi^2$ of the fit improves when a sinusoid of period
$P$ is included (see Cumming 2004 for a more detailed description).

Since analytical approaches tend to give optimistic confidence levels,
it is desirable to obtain the False Alarm Probability of a solution
empirically. To make this, we generate a large number of synthetic
datasets ($10^5$) with the same sampling cadence (same dates) but only
containing random noise. Then, for each realization, we compute the
Least Squares periodogram and find the period with higher power, which
will be a false alarm. A histogram of False Alarms as a function of
the Power is obtained and its complementary cumulative distribution
gives the False Alarm Probability of a given peak in our signal. The
Least Squares periodogram of the RVS data and its associated empirical
FAP probability distribution are shown in
Fig.~\ref{fig:fig4}. Compared to the Lomb-Scargle periodogram, this
approach allows a proper weighting of each observation and can be
easily generalized to include other time-dependent effects in the
signal at the period search level.
\end{appendix}

\end{document}